\documentclass[pra, notitlepage, nofootinbib]{revtex4-1}

\usepackage{geometry}
\usepackage{epsfig}
\usepackage{epstopdf}
\usepackage{bm}
\usepackage[nooneline]{caption}

\newcommand{\ket}[1]{{|{#1}\!\!>}}

\begin{document}

\title{Quantum States as Ordinary Information}

\author{Ken Wharton}
%\email{wharton@science.sjsu.edu}
\affiliation{Department of Physics and Astronomy, San Jos\'{e} State University, San Jos\'{e}, CA 95192-0106}
%\date{September 3, 2007}                                           % Activate to display a given date or no date
\begin{abstract}

Despite various parallels between quantum states and ordinary information, quantum no-go-theorems have convinced many that there is no realistic framework that might underly quantum theory, no reality that quantum states can represent knowledge \textit{about}. This paper develops the case that there \textit{is} a plausible underlying reality: one actual spacetime-based history, although with behavior that appears strange when analyzed dynamically (one time-slice at a time).  By using a simple model with \textit{no} dynamical laws, it becomes evident that this behavior is actually quite natural when analyzed ``all-at-once'' (as in classical action principles).  From this perspective, traditional quantum states would represent incomplete information about possible spacetime histories, conditional on the future measurement geometry.  Without dynamical laws imposing additional restrictions, those histories can have a classical probability distribution, where exactly one history can be said to represent an underlying reality.

\end{abstract}

\maketitle

\setlength{\baselineskip}{1.25\baselineskip} 

\captionsetup[figure]{justification=raggedright}

%%%%%%%%%%%%%%%%%%%%%%%%%%%%%%%%%%%%%%%%%%
\vspace{-12pt}
\section{Introduction}

There are many parallels between quantum states and states of classical knowledge.  A substantial number of quantum phenomena (including entanglement and teleportation) can be shown to have a strong analog in classical systems for which one has restricted knowledge \cite{Spekkens}, and the quantum collapse is strongly reminiscent of Bayesian updating upon learning new information \cite{Fuchs}.  These two well-cited papers alone provide numerous examples and arguments that link quantum information theory to classical information about quantum systems.

However, despite these connections between information/knowledge and quantum states, there has been little progress towards answering the deeply related ``quantum reality problem'' \cite{Kent}: What is the underlying reality that quantum states represent knowledge \textit{about}?  If quantum states are information, what is the ``informata''?  Most research in quantum information has ducked this question, or denied the very possibility of hidden variables, likely due to strong no-go theorems \cite{Bell,KS,PBR}.  Indeed some have rejected the basic notion that information needs to be about anything at all, taking the ``It from Bit'' view that information can somehow be more fundamental than reality. \cite{Wheeler}

This paper will lay out the case that treating the quantum state as a state of knowledge does not require the ``It from Bit" viewpoint.  This will be accomplished by demonstrating that---despite the no-go theorems and conventional wisdom---quantum states can be recast as incomplete classical information about an underlying spacetime-based reality.   Note that the specific meaning of ``information'', as used in this paper, refers exclusively to an agent's knowledge.  Timpson \cite{Timpson} has carefully shown why this meaning is crucially distinct from the technical concept of Shannon Information, which is perhaps better termed ``source compressibility'' or ``channel capacity'' (in different contexts), and is a property of \textit{real} sources or channels.  Given that this paper sets to one side the alternate viewpoint that the quantum state comprises an element of reality (at least until the final conclusion), the conventional use of the word ``information" is most appropriate for this discussion.

The below conclusions do not contradict the quantum no-go theorems.  These theorems have indeed ruled out local hidden \textit{instructions}; there can be no local hidden variables (LHVs) that determine quantum outcomes if the LHVs are independent of the (externally controlled) experimental geometry in that system's future.  Such a ``no retrocausality'' premise is a natural assumption for a system that solves a local dynamical equation, but fails for locally-interacting systems in general.  If the LHVs are not merely instructions, but also reflect future measurement settings, the no-go theorems are not applicable.  This is the often-ignored ``retrocausal loophole'', that has been periodically noted as a way to restore a spacetime-based reality to quantum theory \cite{OCB,CWR,Cramer,Sutherland1,Price,Miller,WMP,Wharton13}.

The main result of this paper is to show how non-dynamical models can naturally resolve the quantum reality problem using the ``all-at-once''-style analysis of action principles (and certain aspects of classical statistical mechanics).  This perspective naturally recasts our supposedly-complete information about quantum systems into \textit{incomplete} information about an underlying, spacetime-based reality.  As this paper is motivating a general research program, rather than a specific answer to the quantum reality problem, the below analysis will strive to be noncommittal as to the precise nature of the informata (or ``beables'') of which quantum states might encode information.  (Although it will be strictly assumed that the beables reside in spacetime; more on this point in Section 2.2.)

After some motivation and background in the next section, a simple model will then demonstrate how the all-at-once perspective works for purely spatial systems (without time).  Then, applying the same perspective to spacetime systems will reveal a framework that can plausibly serve as a realistic account of quantum phenomena.  The result of this analysis will be to dramatically weaken the ``It from Bit'' idea, demonstrating that realistic beables \textit{can} exist in spacetime, no-go theorems not withstanding---so long as one is willing to drop dynamics in favor of an all-at-once analysis.  We may still choose to reject this option, but the mere fact that it is on the table might encourage us \textit{not} to redefine information as fundamental---especially as it becomes clear just how poorly-informed we actually are.

\section{Framework and Background}

\subsection{Newtonian vs. Lagrangian Schemas}

Isaac Newton taught us some powerful and useful mathematics, dubbed it the ``System of the World'', and ever since we've assumed that the universe actually runs according to Newton's overall scheme.  Even though the details have changed, we still basically hold that the universe is a computational mechanism that takes some initial state as an input and generates future states as an output.  

Such a view is so pervasive that only recently has anyone bothered to give it a name: Lee Smolin now calls this style of mathematics the ``Newtonian Schema'' \cite{Smolin}.  Despite the classical-sounding title, this viewpoint is thought to encompass all of modern physics, including quantum theory.  This assumption that we live in a Newtonian Schema Universe (NSU) is so strong that many physicists can't even articulate what other type of universe might be conceptually possible.  

When examined critically, the NSU assumption is exactly the sort of anthropocentric argument that physicists usually shy away from.  It is essentially the assumption that the way we solve physics problems must be the way the universe actually operates.  In the Newtonian Schema, we first map our knowledge of the physical world onto some mathematical state, then use dynamical laws to transform that state into a new state, and finally map the resulting (computed) state back onto the physical world.  This is useful mathematics, because it allows us to predict what we don't know (the future), from what we do know (the past).   But it is possible we have erred by assuming the universe must operate as some corporeal image of our calculations.

The alternative to the NSU is well-developed and well-known: Lagrangian-based action principles.  These are perhaps thought of as more a mathematical trick than as an alternative to dynamical equations, but the fact remains that all of classical physics can be recovered from action-extremization, and Lagrangian Quantum Field Theory is strongly based on these principles as well.  This indicates an alternate way to do physics, without dynamical equations---deserving of the title ``the Lagrangian~Schema''.

Like the Newtonian Schema, the Lagrangian Schema is a mathematical technique for solving physics problems.  One sets up a (reversible) two-way map between physical events and mathematical parameters, partially constrains those parameters on some spacetime boundary \emph{at both the beginning and the end}, and then uses a global rule to find the values of the unconstrained parameters and/or a transition amplitude.  This analysis does not proceed via dynamical equations, but rather is enforced on entire regions of spacetime ``all at once''.

While it's a common claim that these two schemas are equivalent, different parameters are being constrained in the two approaches.  Even if the Lagrangian Schema yields equivalent dynamics to the Newtonian Schema, the fact that one uses different inputs and outputs for the two schemas ({\it i.e.}, the final boundary condition is an input to the Lagrangian Schema) implies they are not exactly equivalent.  And conflating these two schemas simply because they often lead to the same result is missing the point: These are still two different ways to solve problems.  When \emph{new} problems come around, different schemas suggest different approaches.  Tackling every new problem in an NSU (or assuming that there is always a Newtonian Schema equivalent to every possible theory) will therefore miss promising~alternatives.  

Given the difficulties in finding a realistic interpretation of quantum phenomena, it's perhaps worth considering another approach: looking to the Lagrangian Schema not as equivalent mathematics, but as a different framework that can be altered to generate physical theories not available to Newtonian Schema approaches.  At first pass, the Lagrangian Schema does indeed seem to naturally solve various perplexing features of NSU-style quantum theory \cite{FQXi4}, but these are mostly big-picture arguments.  The subsequent sections will indicate how this might work in practice.   

\subsection{Previous Work}

In physical models without dynamical laws, everything must be kinematics.  Instead of summarizing a system as a three-dimensional ``state'' (subject to temporal dynamics), the relevant system is now the entire four-dimensional ``history''.  To solve the quantum reality problem, then, there must be exactly one (fine-grained) history that actually occurs.  Crucially, the history should not be additionally constrained by dynamical laws, or one effectively reverts back to the Newtonian Schema, even for a history-based~analysis.

In quantum foundations, analyzing histories without dynamics is uncommon but certainly not unheard of; several different research programs have pursued this approach.  Still, in seemingly every one of these programs, the history-analysis is accompanied with a substantial modification to (A) ordinary spacetime, or (B) ordinary probability and logic.  Looking at previous research, one might conclude that it is not the Lagrangian Schema analysis that resolves problems in quantum foundations, but instead one of these other dramatic modifications.  But such a conclusion is incorrect; the central point of this paper is that an all-at-once framework can naturally resolve all of the key problems without requiring any changes to (A) or (B).

Before turning to a brief summary of such research programs, it should be noted that any approach with an ontology that solves dynamical equations (such as the standard quantum wavefunction) will fall under the Newtonian Schema and will be subject to the no-go theorems; they will not have a spacetime-based resolution to the quantum reality problem.  Even stochastic deviations from dynamic equations fall under the Newtonian Schema, as the future state is determined from the past state in addition to random inputs.  Such approaches are not the target of this paper, even those that treat solutions to the dynamic equations as entire histories.  (The point of this paper is to show that realistic beables {\em can} exist in spacetime, and that the quantum state {\em can} encode information about these beables; not that all other approaches must be wrong.)  Therefore, any history-based analysis of deterministic Bohmian trajectories and/or stochastic Ghirardi-Rimini-Weber (GRW) theories \cite{Goldstein} are not the sort of theory with which we are presently concerned.  

Furthermore, any ontology built from the standard wavefunction is effectively a dramatic modification to spacetime (A), because multiparticle wavefunctions (or field functionals, in the case of quantum field theory) do not reside in ordinary spacetime.  (They instead reside in a higher-dimensional configuration space.)  Even setting aside standard Bohmian \cite{dBB} and Everettian~\cite {Everett} interpretations of quantum theory (which treat the wave function as ontological, not epistemic), there are other wavefunction-based approaches that (arguably) have some non-dynamical element (including GRW-style flash ontologies~\cite{Bell89,Kent89,Tumulka}, Cramer's Transactional Interpretation \cite{Cramer} and the Aharonov-Vaidman two-state approach \cite{AV}).  Even if these approaches somehow argued that they did not take the \textit{standard} wavefunction to be ontological, their beables are still clearly functions on configuration space, not spacetime (A).

Without the wavefunction as an ontological element, there are still several history-based approaches in the literature.  Griffiths' ``Consistent Histories'' framework \cite{Griffiths} is one example, although it is not a full solution to the quantum reality problem, as there are many cases where no consistent history can be found.  Also, there is never {\em one} fine-grained history that can be said to occur.  Gell-Mann and Hartle have recently \cite{GellMann} attempted to resolve these problems, but in the process they modify probabilistic logic (B), enabling the use of negative probabilities.  (Tellingly, they also demand that the fine grained histories take the form of particle trajectories, even if the future measurement involves path interference.)

Several history-based approaches have been championed by Sorkin and colleagues.  A research program motivated and based upon the path integral \cite{Sorkin} is of particular relevance, although it is almost always presented in the context of a non-classical logic (B).  (Not all such work falls in this category; one notable exception is a recent preprint by Kent \cite{Kent2}).  This path-integral analysis is rather separate from Sorkin's causal set program \cite{Causet}, which seeks to discretize spacetime in a Lorentz-covariant manner.  While this is also history-based, it clearly modifies spacetime (A).   

Another approach by Stuckey and Silberstein (the Relational Blockworld \cite{Stuckey}) is strongly aligned against the Newtonian Schema, and the all-at-once aspect is central to that program.  But again, this history-based framework comes with a severe modification of spacetime (A), in that the Relational Blockworld replaces spacetime with a discrete substructure.  It is therefore unclear to what extent this program resolves interpretational questions via the non-existence of ordinary spacetime rather than simply relying on the features of all-at-once analysis. 

Finally, an interesting approach that maps the standard quantum formalism onto a more time-neutral framework is recent work by Leifer and Spekkens \cite{Leifer}.  Notably, it explicitly allows updating one's description of the past upon learning about future events.  Because the quantum conditional states defined in this work are clearly analogous to states of knowledge rather than states of reality, the fact that they exist in a large configuration space is not problematic (and indeed there is a strong connection to work built on spacetime-based beables \cite{WMP}).  Still, perhaps because of the lack of realistic underlying beables, the logical rules required to extract probabilities from these states differ somewhat from classical probability theory (B).

Any of the above research programs may turn out to be on the right track; after all, there is no guarantee that the beables that govern our universe {\em do} exist in spacetime.  But the fact that they all modify spacetime (or logic) has thoroughly obscured a crucial point: A history-based analysis, with no dynamical restrictions, \textit{need not modify spacetime or logic to resolve the quantum reality problem}, even taking the no-go theorems into account.  The next sections will demonstrate how an ``all at once'' approach can provide a realistic underlying framework for quantum theory, and then will interpret quantum states in this light as encoding ordinary information about underlying spacetime-local beables.

\section{The Partial-Knowledge Ising Model}

Analyzing systems without the use of dynamical equations may be counter-intuitive, but it is done routinely; one can analyze 3D systems for which there are no dynamics, by definition.  A particularly useful approach is found in classical statistical mechanics (CSM), because in that case one never knows the exact microscopic details, providing a well-defined framework in which to calculate probabilities that result from partial knowledge.

A specific example of this CSM-style logic can be found in the classical Ising model.  Specifically, one can imagine agents with partial knowledge about the exact connections of the Ising model system, and then reveal knowledge to those agents in stages.  The resulting probability-updates will be seen to have exactly the features thought to be impossible in dynamical systems.  The fact that this updating is natural in a classical, dynamics-free system such as the Ising model will demonstrate that it is also natural in a dynamics-free (``all at once'') analysis of spacetime systems.  

The classical Ising model considered here is very small to ensure simplicity: 3 or 4 lattice sites labeled by the index $j$, with each site associated with some discrete variable $\sigma_j=\pm 1$.  The total energy of the system is taken to be proportional to
\begin{equation}
\label{eq:Ising}
H(\sigma)=-\sum_{<i\,j>} \sigma_i \sigma_j,
\end{equation}
where the sum $<i\,j>$ is over neighboring lattice sites.  If the system is known to be in equilibrium at an inverse temperature $\beta=(k_B\,T)^{-1}$, the (joint) probability of any complete configuration $\sigma$ is proportional to $exp[-\beta H(\sigma)]$.  These relative probabilities can be transformed into absolute probabilities by the usual normalization procedure, dividing by the partition function
\begin{equation}
\label{eq:Z}
Z=\sum_\sigma e^{-\beta H(\sigma)},
\end{equation}
where the sum is over all allowable configurations.  Therefore, the probability for any specific configuration is given by
\begin{equation}
\label{eq:P}
P(\sigma)=\frac{e^{-\beta H(\sigma)}}{Z}.
\end{equation}

Several obvious features should be stressed here, for later reference.  Most importantly, there is exactly one actual state of the system $\sigma$ at any instant; other configurations are possible, but only one is real.  This implies that (at any instant) the probabilities $P(\sigma)$ are \textit{subjective}, in that they represent degrees of belief but not any feature of reality.  Therefore, one could learn information about the system (for example, that $\sigma_1=1$) that would change $P(\sigma)$ without changing the system itself.  (Indeed, learning such knowledge would also change $Z$, as it would restrict the sum to only global configurations where $\sigma_1=1$.)  Finally, note that the probabilities $P(\sigma)$ do not take the form of local conditional probabilities, but are more naturally viewed as \textit{joint} probabilities over all lattice points that make up the complete configuration $\sigma$.  

The below analysis will work for any finite value of $\beta$, but it is convenient to take $\beta=ln(\sqrt{2})$.  At this temperature, for a minimal system of two connected lattice sites, the lattice values are twice as likely to have the same sign as they are to have opposite signs: $P(\sigma_1\!\!=\!\!+1,\sigma_2\!\!=\!\!+1)=2P(\sigma_1\!\!=\!\!+1,\sigma_2\!\!=\!\!-1)$.
\newpage

\begin{figure}[htb]
\centerline{\includegraphics[width=.45\textwidth]{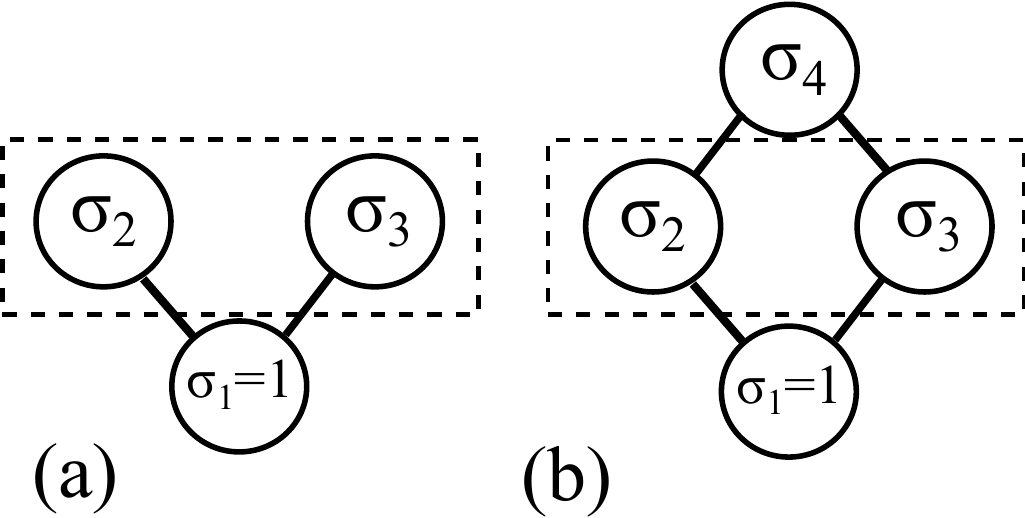}}
\caption{Two geometries of an Ising model; in both cases one knows the bottom lattice site is $\sigma_1=1$.  A dashed box indicates the subsystem of interest.  The central example is where one does not know whether the geometry is that of Figure \ref{Figure:Fig1}a or \ref{Figure:Fig1}b.}
\label{Figure:Fig1}
\end{figure}

For a simple lattice that will prove particularly relevant to quantum theory, consider Figure \ref{Figure:Fig1}a.  Each circle is a lattice site, and the lines show which sites are adjacent.  With only three lattice sites, and knowledge that $\sigma_1=1$, it is a simple manner to calculate the probabilities $P(\sigma_2,\sigma_3)$.  For $\beta=ln(\sqrt{2})$, the above equations yield probabilities:
\begin{center}
\begin{tabular}{cc|c}
$\sigma_2$  & $\sigma_3$  & $P_{1a}(\sigma_2,\sigma_3)$ \\
\hline
$+1$ & $+1$ & $4/9$ \\
$+1$ & $-1$ & $2/9$ \\
$-1$ & $+1$ & $2/9$ \\
$-1$ & $-1$ & $1/9$ \\
\end{tabular}
\end{center}

For Figure 1b, the probabilities require a slightly more involved calculation because the fourth lattice site changes the global geometry (from a 1D to a 2D Ising model).  In this case, the joint probability $P(\sigma_2,\sigma_3)$ can be found by first calculating $P(\sigma_2,\sigma_3,\sigma_4)$ and then summing over both possibilities \mbox{$\sigma_4=\pm 1$}.  This process yields different probabilities:
\begin{center}
\begin{tabular}{cc|c}
$\sigma_2$  & $\sigma_3$  & $P_{1b}(\sigma_2,\sigma_3)$ \\
\hline
$+1$ & $+1$ & $20/41$ \\
$+1$ & $-1$ & $8/41$ \\
$-1$ & $+1$ & $8/41$ \\
$-1$ & $-1$ & $5/41$ \\
\end{tabular}
\end{center}

The most interesting example is a further restriction where an agent \textit{does not know} whether the actual geometry is that of Figure \ref{Figure:Fig1}a or \ref{Figure:Fig1}b.  Specifically, one knows that $\sigma_1=1$, and that $\sigma_2$ and $\sigma_3$ are both adjacent to $\sigma_1$.  But one does not know whether $\sigma_4$ is also adjacent to $\sigma_2$ and $\sigma_3$ ({Figure 1b}), or whether $\sigma_4$ is not even a lattice site ({Figure 1a}).

This is not to say there is no fact of the matter; this example presumes that \textit{is} some particular geometry---it is merely unknown.  This is not quite the same as the unknown values of a lattice site (which also have some particular state at any given instant), because the Ising model provides no clues as to how to calculate the probability of a \textit{geometry}.  All allowable states may have a known value of $exp[-\beta H(\sigma)]$ from Equation (\ref{eq:Ising}), but without knowledge of which states are allowable one cannot calculate $Z$ from Equation (\ref{eq:Z}), and therefore one cannot calculate probabilities in the configuration space $P(\sigma_2,\sigma_3)$. 

The obvious solution to such a dilemma is to use a configuration space conditional on the unknown geometry $G=1a$ or $G=1b$, assigning probabilities to $P(\sigma_2,\sigma_3|G)$.  Indeed, this has already been done in the above analysis using the notation $P_{G}(\sigma_2,\sigma_3)$.  Note that one cannot include $G$ in the configuration space as $P(\sigma_2,\sigma_3,G)$ because (unlike $P_G$) such a distribution would have to be normalized, and there is no information as to how to apportion the probabilities between the $G=1a$ and $G=1b$ cases.

Given this natural and (presumably) unobjectionable response to such a lack of knowledge, the next section will explore a crucial mistake that would lead one to conclude that \textit{no underlying reality exists} for this CSM-based model, despite the fact that an underlying reality does indeed exist (by construction).  Then, by applying the above logic to a dynamics-free scenario in space \textit{and} time, one can find a strong analogy to probabilities in quantum theory with an unknown future measurement.  

\section{Implications of the Model}\vspace{-12pt}
\subsection{The Independence Fallacy}

In the above partial-knowledge Ising model, the subsystem of interest is the dashed box that comprises the lattice sites $\sigma_2$ and $\sigma_3$.  (In any physical system, one might reasonably want to analyze a portion of it, independent from the rest.)  The cost of this analysis, however, is that the mathematical structure over which a (limited knowledge) agent can ascribe probabilities is the conditional configuration space $P_G(\sigma_2,\sigma_3)$ rather than a ``pure'' configuration space over all possibilities $P(\sigma_2,\sigma_3)$. 

It should be clear that a serious fallacy would result if one were to demand that probabilities should be assigned to such a pure configuration space.  Such a demand would be natural if one thought that the probabilities ascribed to the subsystem should be independent of the external geometry, but would lead to the false equation $P_{1a}=P_{1b}$.  This mistake will be termed the ``Independence Fallacy''.  From the above analysis, the explicitly calculated $P_{1a}$ and $P_{1b}$ reveals that they are not the same, and therefore this fallacy would lead to contradictions.  

It is useful to imagine a confused agent who for some reason tried to build a theory around the (mistaken) premise that one should always be able to ascribe a master probability distribution $P(\sigma_2,\sigma_3)$ that was independent of geometry.  Such an agent would quickly find that this was an impossible task, and that there was no such classical probability distribution over the realistic microstates of this subsystem.  Given the Independence Fallacy, the analysis of the confused agent might go like this:  The geometry of Figure 1a implies a 5/9 probability of $\sigma_2=\sigma_3$, while the geometry of Figure {1b} implies a 25/41 probability.  But since this value must be independent of the geometry, the question ``Does $\sigma_2=\sigma_3$?'' cannot be assigned a coherent probability.  And if it cannot be answered, such a question should not even be asked.

This, of course, is wrong: such a question \textit{can} be asked in this model, but the answer depends on the geometry.  It is the Independence Fallacy which might lead to a denial of an underlying reality, and the ultimate culprit (in this case, at least) would be the attempt to describe a subsystem independently from the entire system.

\subsection{Information-Based Updating}

Without the Independence Fallacy, it is obvious how the conditional probabilities $P(\sigma_2,\sigma_3|G)$ should be updated upon learning new information.  For example, if an agent learned that the geometry was in fact that of Figure \ref{Figure:Fig1}b, a properly-updated description would simply be $P_{1b}$.  There would be no reason to retain $P_{1a}$ as a partial description of the system; it would merely represent probabilities of a counter-factual geometry.  Discarding this information is clearly Bayesian updating, not a physical change in the system.  (Further updating would occur upon learning the actual value of a lattice site.)

The central point is that some information-updating naturally occurs when one learns the geometry of the model, even without any revealed lattice sites.  And because CSM is a realistic model (with some real, underlying state), the information updating has no corresponding feature in objective reality.  Updating is a subjective process, performed as some agent gains new information.

\subsection{Introducing Time}

The above Ising model was defined as a static system in two spatial dimensions.  The only place that time entered the analysis was in the updating process in the previous subsection, but this subjective updating was not truly ``temporal'', as it had no relation to any time in the system.  Indeed, one could give different agents information in a different logical order, leading to different updating.  Both orders would be unrelated to any temporal evolution of the system (which is static by definition).

Still, an objective time coordinate can be introduced in a trivial manner: simply redefine the model such that one of the spatial axes in Figure \ref{Figure:Fig1} represents time instead of space.  Specifically, suppose that the vertical axis is time (past on the bottom, future on the top).  It is crucial not to introduce dynamics along with time; one point of the model was to show how to analyze systems without dynamics.  This analysis has already been performed, and does not change.  The dashed box in  Figure \ref{Figure:Fig1} now represents a space-time subsystem (\textit{i.e.}, one instant in time), and the same state-counting logic will lead to exactly the same probabilities as the purely spatial case.

One might be tempted to propose reasons why this space-time model is fundamentally different from the original space-space model, perhaps assuming the existence of dynamical laws.  Such laws \textit{would} break the analogy, but they are not part of the model.  One might also argue that the use of $H$ in Equation~(\ref{eq:Ising}) embedded a notion of time even in the spatial case, but Equation (\ref{eq:P}) can be written entirely in terms of the unitless parameters $\beta$ and $\sigma_j$, not requiring any notion of a Hamiltonian.   After all, the equations of CSM result from assigning equal \textit{a priori} probabilities to all possible microstates; one can get similar equations in a spacetime framework by simply assigning equal \textit{a priori} probabilities to all possible temporally-extended microstates (or more intuitively, ``microhistories'') \cite{Wharton13}.

Whether or not the analogy \textit{can} be broken, the previous section is an existence proof that such a system \textit{could} be analyzed in this manner, which is all that is needed for the below conclusions.  It is logically possible to assign a relative probability to each spacetime-history, normalize these probabilities using the entire allowed space of microhistories,  and then make associated predictions.  If one did so, it would be natural to update one's probabilities of an instantaneous subsystem when one learned about the experimental geometry in that subsystem's future.

The unusual features of the above partial-knowledge Ising model should now have a clear motivation.  For an agent not to know the spatial geometry (Figure 1a vs. Figure 1b) would typically be an artificial restriction.  But it is quite natural not to know the future, and if the vertical axis represents time, it is more plausible that an agent might be uncertain as to whether $\sigma_4$ would exist.  However, this does not break the analogy, either.  While we tend to learn about things in temporal order, it's not a formal requirement; we can film a movie of a system and analyze it backwards, or even have spatial slices of a system delivered to us one at a time.  The link between information-order and temporal-order is merely typical, not a logical~necessity.

In a space-time context, it is also more understandable how one might fall into the Independence Fallacy in the first place.  If we expect the future to be generated from the past via the Newtonian Schema, then we would also expect the probabilities we assign to the past to be independent of the future experimental geometry.  But without dynamical laws, this is provably incorrect.  If we assign every microhistory a probability, the conditional probabilities $P(\sigma_2,\sigma_3|G)$ that made sense in the spatial case also make sense in the temporal case.  When we learn about the experimental geometry \textit{of the future}, such an analysis would update our probabilistic assessment of the past.

\section{Quantum Reality}

From the perspective of the above analysis, as applied to systems in both space and time, the various quantum no-go theorems are effectively making the Independence Fallacy.  Specifically, they assume that any probabilistic description of the present must be independent of the future experimental geometry.  In some discussions this assumption is not even explicit, with authors implicitly assuming that one \textit{obviously} should not update one's (past) probabilities upon learning about the future measurement setting~\cite{Maudlin}.

However, what may seem (to some) to be obvious in a space-time setting is provably wrong in the above space-space setting, where the probabilities of a subsystem do depend on the external geometry.  It is also provably wrong in a space-time setting of connected lattice points where the probabilities are given by Equation (\ref{eq:P}) (as opposed to dynamical laws).  Therefore it should at least give one pause to make such an independence assumption in general.  Arguing that standard quantum mechanics is governed by dynamical time evolution does not void this analysis; research programs that attempt to solve the quantum reality problem are exploring \textit{alternatives}, and those alternatives need not in general utilize dynamical laws (as in some examples from Section 2.2).
\vspace{-12pt}
\subsection {Double Slit Experiment}

\begin{figure}[htb]
\centerline{\includegraphics[width=.5\textwidth]{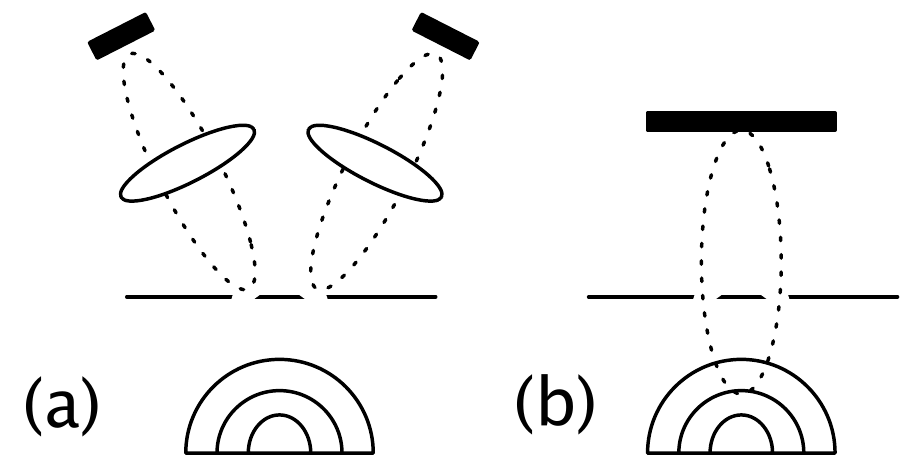}}
\caption{Two geometries of a double slit experiment, in which a single photon passes through a pair of slits.  (The vertical axis is performing double-duty as both time and a second spatial axis.) ({\textbf a}) Lenses and (black) detectors measure which slit the photon passes through; ({\textbf b}) A screen records a photon that contributes to a two-slit interference pattern.}
\label{Figure:Fig2}
\end{figure}

To begin with a simple example, consider the delayed-choice double-slit experiment, depicted in Figure \ref{Figure:Fig2}.  Here a source (at the bottom) generates a single photon that passes up through a pair of slits.  When lenses are used to form an image of the slits on a detector (as in Figure \ref{Figure:Fig2}a), one always finds that the photon passed through one slit \textit{or} the other.  

However, it appears that each photon passes through both slits if one considers the experiment in Figure~2b.  Here a screen records the interference pattern produced by waves passing through \textit{both} slits, built up one photon at a time.  In the many-photon limit, this pattern is explicable \textit{only} if a wave passes through both slits and interferes.  Since each individual photon conforms to this pattern (not landing in dark fringes), the most obvious conclusion is that each photon somehow passes through both slits.

Where reality seems to fail here is the description of the photon at the slits---one instant of the full spacetime diagram.  In Figure \ref{Figure:Fig2}a the photon seems to go through only one slit; in Figure \ref{Figure:Fig2}b it seems to go through both.  And since the status of the photon at the slits is ``obviously'' independent of the future experimental geometry, it follows that the actual location(s) of the photon-wave at the slits cannot be assigned a coherent probability.  

Except that this is \textit{exactly} the Independence Fallacy!  Compare Figure \ref{Figure:Fig1} to Figure \ref{Figure:Fig2}; they are quite analogous.  In Figure 1a and Figure 2a the right and left branches stay separate; in Figure 1b and Figure 2b the geometry begins in the same way, but then allows recombination.  Following the above logic, \textit{avoiding} the Independence Fallacy allows a coherent underlying reality for the double-slit experiment.  

While a generic realistic theory may have a quite complicated description of what is happening at the slits, the key parameter for our purposes is $N$, the number of slits that any beables pass through (classical EM fields, etc.; the details are irrelevant so long as one avoids a particle ontology for which $N\!\!=\!\!1$ by definition).  In a realistic model, it must be that $N\!\!=\!\!1$ or $N\!\!=\!\!2$ in any given case, although since this value is hidden from an external agent, the agent will have a probability distribution $P(N)$ over these two possible values.

Utilizing the above analysis, the comparison to the partial-knowledge Ising model reveals a natural solution to the problem of defining $P(N)$.  The answer is that the appropriate quantity for an agent who does not know the future experimental geometry $G$ is $P(N|G)$ (or equivalently $P_G(N)$).  Specifically, in this case we have $P_{2a}(N\!\!=\!\!1)=P_{2b}(N\!\!=\!\!2)=1$ and $P_{2a}(N\!\!=\!\!2)=P_{2b}(N\!\!=\!\!1)=0$.  Upon learning the future geometry, an agent would update her assessment of the past probabilities, as above.  In this case, $N$ becomes certain, and one has a realistic description of the experiment.  (For 2b, it is perfectly realistic to have a wave go through both slits.)  If one avoids the Independence Fallacy and does not try to enforce $P_{2a}=P_{2b}$, there are no logical problems with this analysis.

\vspace{-12pt}
\subsection {Bell Inequality Violations}

When analyzing experiments that exhibit Bell-inequality violations, the Independence Fallacy takes the form of ``Measurement Independence'' \cite{Shimony}, where one assumes that the probability distribution over any (past) hidden variables $\lambda$ must be independent of future measurement settings.  But measurement settings are just a type of geometry (consider the difference between Figure \ref{Figure:Fig2}a,b), so this assumption takes the form of $P(\lambda|G)=P(\lambda|G')$ for all possible future measurement geometries $(G,G')$, and this is indeed the above-discussed Independence Fallacy. 

\begin{figure}[htb]
\centerline{\includegraphics[width=.65\textwidth]{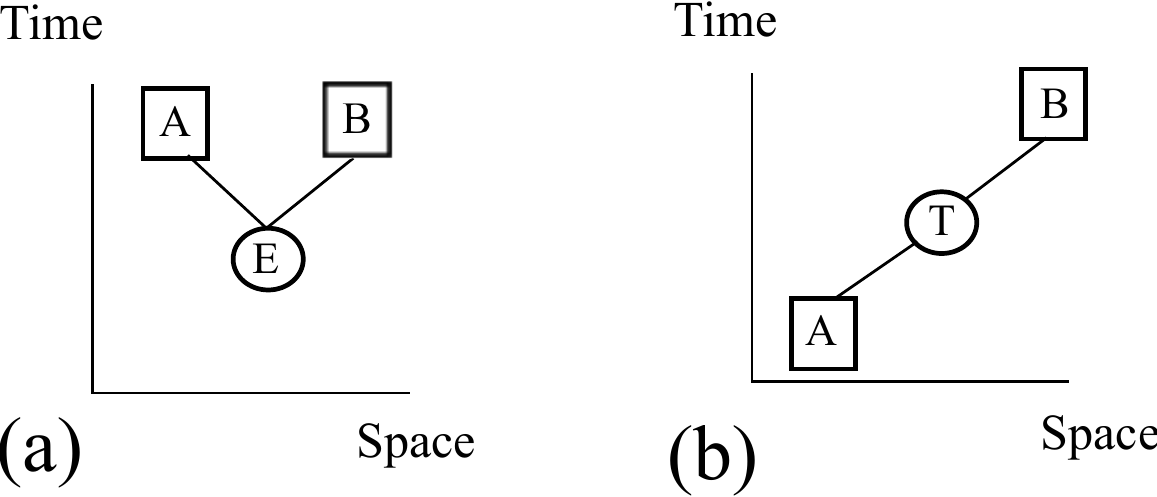}}
\caption{Two dual experiments, with measurements by Alice (A) and Bob (B).  For known measurement settings, there is always a duality between any entangled source (E) and some intermediate operation (T).}
\label{Figure:Fig3}
\end{figure}

Even without committing to a particular ontology, it's possible to see how the all-at-once approach resolves entanglement-based no-go theorems (assuming a spacetime-based ontology that at least locally obeys $T$-symmetry).  Consider the two different experimental geometries shown in Figure \ref{Figure:Fig3}.  In Figure \ref{Figure:Fig3}a, a source $E$ prepares an entangled pair of particles, sends one to Alice ($A$), and one to Bob ($B$); both Alice and Bob freely choose measurements to perform on the particles they receive.  In Figure \mbox{\ref{Figure:Fig3}b}, Alice prepares a single particle; that particle is transmitted to a device $T$ that performs some particular transformation on the particle, and the transformed output is then sent to Bob who performs a final measurement.

Given the stark differences between these two experiments the standard quantum framework, it may come as a surprise that for any entanglement process $E$, there is always a corresponding transformation $T$ that will make the predicted correlations between Alice and Bob identical (between Figure \ref{Figure:Fig3}a and Figure \ref{Figure:Fig3}b).  This duality has been worked out in detail \cite{WMP} (also see \cite{Leifer,Leifer2,EPW}), and the fact that the the connection between $E$ and $T$ requires knowledge of the measurement settings (\textit{i.e.}, the observables at $A$ and $B$) should not be surprising in light of the above arguments.  (In one particularly simple example, if $E$ prepares two spin-1/2 particles in the entangled state $(|00\!>+|11\!>)/\sqrt{2}$, and $A$ and $B$ are both spin measurements in the x-z plane, the corresponding transformation $T$ is merely the identity.) 

This duality between Figure \ref{Figure:Fig3}a and Figure \ref{Figure:Fig3}b allows one to transform any all-at-once account of a single-particle experiment (Figure \ref{Figure:Fig3}b) to an equivalent all-at-once account of an experiment with entanglement (Figure \ref{Figure:Fig3}a).  All that needs to be done is to time-reverse the left half of the Figure \ref{Figure:Fig3}b ontology (which can always be done for a realistic history if the beables exist in spacetime).   Crucially, the all-at-once analysis applies to the entire history, so while a dynamical explanation would not survive this transformation, an all-at-once analysis does.  Unless a given no-go theorem applies to single-particle experiments (and the Bell inequality violations do not), then once can use this duality to turn any time-neutral LHV explanation of a single-particle experiment into a corresponding time-neutral LHV explanation of the entangled dual experiment \cite{WMP}.
 
As an illustrative example, consider Schulman's ansatz that the Bloch-sphere vector of a spin-1/2 system might undergo an anomalous rotation of a net angle $\alpha$ with a probability proportional to $W(\alpha)=(\alpha^2+\gamma^2)^{-1}$ \cite{Schulman}.  (Here $\gamma$ is a small arbitrary angle.)  Consider the case of Figure \ref{Figure:Fig3}b with $T$ as the identity transformation; Alice and Bob simply make consecutive spin measurements on a single spin-1/2 particle, each in some particular direction in the x-z plane.  In accordance with the Lagrangian Schema, it is consistent to have both the preparation and the measurement act as external boundary conditions that force the spin-state to be aligned with the angle chosen by the measuring apparatus (or possibly anti-aligned, in the case of measurement).  Then, for an angle $\theta$ between Alice and Bob's settings, the net anomalous rotation between measurements must be either $\theta$ or $\pi+\theta$, corresponding to Bob's two possible outcomes.  By considering all possible histories with those anomalous rotations (including rotations larger than $2\pi$) this implies
\begin{equation}
\label{eq:Sch1}
\frac{P(\theta)}{P(\pi-\theta)}=\frac{ \displaystyle\sum_{n=-\infty}^{\infty}W(2n\pi+\theta)}{\displaystyle\sum_{n=-\infty}^{\infty}W(2n\pi+\pi+\theta)}=\frac{cos^2(\theta/2)+sin^2(\theta/2)tanh^2(\gamma/2)}{sin^2(\theta/2)+cos^2(\theta/2)tanh^2(\gamma/2)}.
\end{equation}
In the limit that $\gamma\to 0$, this obviously reduces to the standard Born rule probabilities, and since $\gamma$ is an arbitrary parameter, Schulman's ansatz can be made arbitrarily close to the Born rule.  (For an expanded analysis, see \cite{Wharton13}).  

Schulman did not apply his ansatz to cases of entanglement, but using the above duality one can easily generate a realistic model of a Bell-inequality violating experiment.  Switching to Figure \ref{Figure:Fig3}a, the equivalent entangled state $E$ predicted by standard quantum mechanics is $(|00\!>+|11\!>)/\sqrt{2}$, but this is a nonlocal state, with no spacetime representation.  Instead, using the duality, one need only utilize a single spin state for each particle, classically correlated at the source $E$ but nowhere else.  (Since $T$ was the identity, the duality merely requires the two particles to be in the same state at $E$.)  The previous ontology, of a spin state undergoing a net anomalous rotation somewhere between Alice and Bob, follows through from Figure \ref{Figure:Fig3}b to \ref{Figure:Fig3}a; there is now an anomalous rotation continuously distributed across \textit{both} particles, of a net angle $\theta$ between Alice's and Bob's measured spin directions.  The relative probability of such a rotation is the same as before, $W(\theta)$.

The rest of the analysis is straightforward.  Alice has a spin measurement setting at some angle $\alpha$ (in the x-z plane), and Bob has a spin measurement setting at some angle $\beta$ (in the x-z plane).  The two possible outcomes for Bob (constrained as before) are now $\beta$ ($b=1$) or $\pi+\beta$ ($b=2$).  Alice has lost some control compared to Figure \ref{Figure:Fig3}b (see \cite{PriceWharton} for details), but the same constraints as Bob implies that her spin-state will be measured at $\alpha$ ($a=1$) or $\pi+\alpha$ ($a=2$).  The correlation between Alice and Bob's outcomes can then be calculated according to
\begin{equation}
\label{eq:corr}
P(a=b)-P(a\ne b)=\frac{ \displaystyle\sum_{n=-\infty}^{\infty}W(2n\pi+\alpha-\beta)-\displaystyle\sum_{n=-\infty}^{\infty}W(2n\pi+\pi+\alpha-\beta)}{\displaystyle\sum_{n=-\infty}^{\infty}W(2n\pi+\alpha-\beta)+\displaystyle\sum_{n=-\infty}^{\infty}W(2n\pi+\pi+\alpha-\beta)}
\end{equation}
In the $\gamma\to 0$ limit this becomes simply $cos(\alpha-\beta)$, which violates the Bell inequalities exactly as would be expected from the traditional entangled state.  

Notice that this resolution still permits Alice and Bob to freely choose their measurement settings.  Their decision lies outside the system considered in Figure \ref{Figure:Fig3}, and their choices are effectively external (final) boundaries on the system of interest.  The LHVs, on the other hand, are in the system, and constrained by the solution space of possible histories.  This (retrocausal) resolution is therefore strikingly distinct from ``superdeterminism'', an alternate conspiratorial approach that requires Alice's and Bob's decisions to somehow be caused by past hidden variables (presumably via Newtonian Schema processes).  Experimental proposals to rule out superdeterminism \cite{SDPRL} therefore have little to say about Lagrangian-Schema retrocausal models, where the past hidden variables are \textit{direct consequences} of future constraints, rather than circuitous causes of future decisions. 

\subsection{Other Theorems}

Most other no-go theorems work in the same manner as above; as very few of them apply to non-entangled states, there is no barrier to a realistic all-at-once account that resolves the dual experiment of Figure \ref{Figure:Fig3}b.  Such an account can then always be translated into the entangled case of Figure \ref{Figure:Fig3}a, so long as the future measurement settings are known and accounted for.  The most notable case of a no-go theorem applying to a non-entangled state is that of Kochen-Specker contextuality \cite{KS}.  But this theorem merely proves that for a realistic account of quantum mechanics, different measurement procedures must imply a different preceding state of reality.  In other words, Kochen-Specker contextuality \textit{is} the failure of the Independence Assumption.  (In the example in the previous section, a different future measurement setting would indeed correspond to a different spin state immediately before measurement.)

Another interesting case is the recent Pusey-Barrett-Rudolph (PBR) theorem \cite{PBR}, which has no uncertainty about the future measurement setting, but explicitly assumes that knowledge of this setting has not led to any updating of an initial hidden state.  One of the two key premises of the PBR theorem is that the initial states of the two systems are completely uncorrelated, even at the level of hidden variables.  But we've seen in the case of the double slit experiment that once an agent updates on a future measurement setting, this can inform details of a correlation in the past, even for a spatially-distributed beable like an electric field.  Another way to think about PBR is as the time-reverse of a typical entanglement experiment.  Since state-counting is explicitly a time-neutral analysis, past correlations can arise in PBR for exactly the same reason that future correlations arise in entanglement experiments.

\section{Lessons for the Quantum State}

While the analysis in the preceding sections can inform a realistic interpretation of quantum theory, it also implies that assigning probabilities to instantaneous configuration spaces is not the most natural perspective.  The above CSM-style analysis is performed on the entire system ``at once'', rather than one time-slice at a time, in keeping with the spirit of the Lagrangian Schema discussed in Section~2.  The implication is that the best way to understand quantum phenomena is not to think in terms of instantaneous states, but rather in terms of entire histories.

Restricting one's attention to instantaneous subsystems is necessary if one would like to compare the above story to that of traditional quantum theory, but it is certainly not necessary in general.  Indeed, general relativity tells us that global foliations of spacetime regions into spacelike hypersurfaces may not even always be possible or meaningful.  Moreover, if the foliation is subjective, the objective reality lies in the 4D spacetime block.  It is here where the histories reside, and to be realistic, one particular microhistory must \textit{really be there}---say, some multicomponent covariant field $\mu(\bm{x},t)$.  A physics experiment is then about \textit{learning} which microhistory $\mu$ actually occurs, via information-based updating; we gain relevant information upon preparation, measurement setting, and measurement itself.  

Given this notion of an underlying reality, one can then work backwards and deduce what the traditional quantum state must correspond to.  Consider a typical experimenter as an agent informed about a system's preparation, but uninformed about that system's eventual measurement.  With no information about the future measurement geometry, the agent would like to keep track of all possible eventual measurements, or $P(\mu(\bm{x},t)|G)$ for all possible values of $G$ (including all possible times when measurement(s) will be made).  This must correspond to the quantum state, $\ket{\psi(t)}$.  

We can check this result in a variety of ways.  First of all, as the number of possible measurements increases, the dimensionality of the quantum state must naturally increase.  This is correct, in that as particle number increases, the number of experimental options also increases (one can perform different measurements on each particle, and joint measurements as well).  This increase in the space of $G$ with particle number therefore explains the increase of the dimensionality of Hilbert space with particle~number.  

This perspective also naturally explains the reduction of the Hilbert space dimension for the case of identical particles.  For two distinguishable particles, the experimenter can make a choice between (A,B)---measurement A on particle 1 and measurement B on particle 2---or (B,A), the opposite.  But for two identical particles, the space of $G$ (and the experimenter's freedom) is reduced; there is no longer a difference between (A,B) and (B,A).  And with a smaller space of $G$, the quantum state need not contain as much information, even if it encodes all possible future measurements.

Further insights can be gained by considering the (standard, apparent) wavefunction collapse when a measurement is finally made.  Even if the experimenter refuses to update their prior probabilities upon learning the choice of measurement setting (the traditional mistake, motivated by the Independence Assumption), it would be even more incoherent for that agent to not update upon gaining direct new information about $\mu$ itself.  So we \textit{do} update our description of the system upon measurement, and apply a discontinuity to  $\ket{\psi(t)}$.

However, through the above analysis, it should be clear that no physically discontinuous event need correspond to the collapse of $\psi$.  After all, when one learns about the value of a lattice site in the Ising model, one simply updates one's assignment of configuration-space probabilities; no one would suggest that this change of $P(\sigma)$ corresponds to a discontinuous change of the classical system.  (The system might certainly be altered by the measurement procedure, but in a continuous manner, not as a sudden collapse.)  Similarly, for a history-counting derivation of probabilities in a spacetime system, learning about an outcome would lead to an analogous updating.  This picture is therefore consistent with the epistemic view of collapse propounded by Spekkens and others \cite{Spekkens, Fuchs}.  

From this perspective, the quantum state represents a collection of (classical) information that a limited-knowledge agent has about a spacetime system at any one time.  \textit{And it is not our best-possible description.}  In many cases we actually do know what measurement will be made on the system, or at least a vast reduction in the space of possible measurements.  This indicates that a lower-dimensional quantum state is typically available.  Indeed, quantum chemists are already aware of this, as they have developed techniques to compute the energies of complex quantum systems without using the full dimensionality of the Hilbert space implied by a strict reading of quantum physics (e.g., density functional theory).  

If we could determine exactly how to write $\ket{\psi(t)}$ in terms of $P(\mu(\bm{x},t)|G)$, then it would be perfectly natural to update \textit{twice}: once upon learning the type of measurement $G$, and then again upon learning the outcome.  It seems evident that this two-step process would make the collapse seem far less unsettling, as the first update would reduce the possibilities to those that could exist in spacetime.  For example, in the which-slit experiment depicted in Figure \ref{Figure:Fig2}a, if one's possibility-space of histories had already been reduced to a photon on one detector \textit{or} a photon on the other detector, there would be no need to talk about superluminal influences upon measurement.

\section{Conclusions}

Before quantum experiments were ever performed, it may have seemed possible that the Independence Fallacy would have turned out to be harmless.  In classical scenarios, dynamic laws do seem to accurately describe our universe, and refusing to update the past based on knowledge of future experimental geometry does seem to be justified.  And yet, in the quantum case, this assumption prevents one from ascribing probability distributions to everything that happens when we don't look.  If the ultimate source of the Independence Fallacy lies in the dynamical laws of the Newtonian Schema, then before giving up on a spacetime-based reality, we should perhaps first consider the Lagrangian Schema as an alternative.  The above analysis demonstrates that this alternative looks quite promising.

Still, the qualitative analogy between the Ising model and the double slit experiment can only be pushed so far.  And one can go \textit{too} far in the no-dynamics direction: considering \textit{all} histories, as in the path integral, would lead to the conclusion that the future would be almost completely uncorrelated with the past, contradicting macroscopic observations.  But with the above framework in mind, and the path integral as a starting point, there are promising quantitative approaches.  One intriguing option~\cite{Wharton13} is to limit the space of possible spacetime histories (perhaps those for which the total Lagrangian density is always zero).  Such a limitation can be seen to cluster the remaining histories around classical (action-extremized) solutions, recovering Euler-Lagrange dynamics as a general guideline in the many-particle limit.  Better yet, for at least one model, Schulman's ansatz from Section~5 can be derived from a history-counting approach on an arbitrary spin state.  And as with any deeper-level theory that purports to explain higher-level behavior, intriguing new predictions are also indicated. 

Even before the technical project of developing such models is complete, the above analysis strongly indicates that one can have a spacetime-based reality to underly and explain quantum theory.  If this is the case, the ``It from Bit'' idea would lose its strongest arguments; quantum information could be about something real rather than the other way around.  It is true that this approach requires one to give up the intuitive NSU story of dynamical state evolution, so one may still choose to cling to dynamics, voiding the above analysis.  But to this author, at least, losing a spacetime-based reality to dynamics does not seem like a fair trade-off.  After all, if maintaining dynamics as a deep principle leads one into some nebulous ``informational immaterialism'' \cite{Timpson}, one gives up on any reality on which dynamics might operate in the first place.  

More common than ``It from Bit'' is the view that maintains the Independence Fallacy by elevating configuration space into some new high-dimensional reality in its own right \cite{Everett}.  Ironically, that resulting state is \textit{interdependent} across the entire 3N-dimensional configuration space, with the lone exception that it remains independent of its future.   While this is certainly a possibility, the alternative proposed here is simply to link everything together in standard 4D spacetime, with full histories as the most natural entities on which to assign probability distributions.  So long as one does not additionally impose dynamical laws, there is no theorem that one of these histories cannot be real.  Furthermore, by analyzing entire 4D histories rather than states in a 3N-dimensional configuration space, one seems less likely to make conceptual mistakes based on pre-relativistic intuitions wherein time and space are completely~distinct.  

The central lesson of the above analysis is that---whether or not one \textit{wants} to give up NSU-style dynamics---action principles provide us with a logical framework in which we \textit{can} give up dynamics.  The promise, if we do so, is a solution to the quantum reality problem \cite{Kent}, where quantum states can be ordinary information about something that really does happen in spacetime.  This would not only motivate better descriptions of what is happening when we don't look (incorporating our knowledge of future measurements), but would formally put quantum theory on the same dimensional footing as classical general relativity.  After nearly a century of efforts to quantize gravity, perhaps it is time to acknowledge the possibility of an alternate strategy: casting the quantum state as \textit{information} about a generally-covariant, spacetime-based reality.

%%%%%%%%%%%%%%%%%%%%%%%%%%%%%%%%%%%%%%%%%%

\acknowledgements{}

The author would like to thank the Foundational Questions Institute (fqxi.org) for sponsoring their annual essay contests, and awarding third prize to the two essays \cite{FQXi4,FQXi5} that formed the basis of this~paper.

%%%%%%%%%%%%%%%%%%%%%%%%%%%%%%%%%%%%%%%%%%

%=================================================================
% References: Variant A
%=================================================================
% Back Matter (References and Notes)
%----------------------------------------------------------
% Style and layout of the references
\bibliographystyle{mdpi}
\makeatletter
\renewcommand\@biblabel[1]{#1. }
\makeatother

%=================================================================
% References:  Variant B
%=================================================================
% Use the following option to include external BibTeX files:
%\bibliography{lite}
%\bibliographystyle{mdpi}

\end{document}